\documentstyle[12pt,bezier]{article}

              \def\be{\begin{equation}}
              \def\ee{\end{equation}}
              \def\bea{\begin{eqnarray}}
              \def\eea{\end{eqnarray}}
        \def\ba{$$\begin{array}}
        \def\ea{\end{array}$$}
\begin{document}

\begin{center}
{\Large{\bf{ Quantum Renormalization Group  \\
for  \\
1 Dimensional Fermion Systems : \\
a Modified Scheme }}}
\vskip .5 cm   
{{ {\bf A. Langari} \footnote {E-mail : LANGARI@netware2.ipm.ac.ir} 
\hspace{0.2cm} , \hspace{0.2cm} {\bf V. Karimipour} }}
\vskip .5cm
{\it \small {
Department of Physics, 
Sharif University of Technology,\\P.O.Box:11365-9161, Tehran, Iran. \\ 
Institute for Studies in Theoretical Physics
and Mathematics (IPM),
\\P.O.Box: 19395-5531, Tehran, Iran.  }}

\end{center}
\vskip .1cm
\begin{abstract}
Inspired by the superblock method of White, we introduce
a simple modification of the standard Renormalization Group (RG) technique
for the study of quantum lattice systems.
Our method which takes into account the effect of Boundary Conditions(BC),
may be regarded as a simple
way for obtaining first estimates of many properties of quantum lattice systems.
By applying this method
to the 1-dimensional free and interacting fermion system, we obtain the ground state 
energy with much higher accuracy than the standard RG.
We also calculate the density-density correlation function in the 
free-fermion case which shows good agreement with the exact result.

PACS : 71.10.Fd, 71.10.Pm, 05.30.Fk \\
Keywords : Quantum renormalization group, Fermion systems.
\end{abstract} 
\newpage
\def\bra#1{{\langle #1 \vert}}
\def\ket#1{{\vert #1 \rangle}}
\def\x#1{{\sigma_{#1}^{x}}}
\def\z#1{{\sigma_{#1}^{z}}}
\def\y#1{{\sigma_{#1}^{y}}}
\def\d{\Delta}
\def\a{\alpha}
\def\b{\beta}
\def\ge{\ground state energy}
\def\ep{\varepsilon}
\def\c#1{{c_{#1}}}
\def\cd#1{{c_{#1}^{\dagger}}}
\def\n#1{{n_{#1}}}

\section{Introduction}  
In spite of the enormous recent interest in correlated fermion systems, it 
still remains a challenging task to solve the interacting many fermion 
problem. Due to the difficulty of obtaining exact solutions, one often
looks for a reasonably good approximation scheme for solving such problems.
However each approximation scheme has its own limits and draw-backs.
Mean-field methods neglect the effects of fluctuations
to a great extent; the perturbative methods, on the other hand, do not work 
over the whole range of the parameter space. Among the other methods a real 
space renormalization group(RG) technique [1-7], often referred to as the block
RG, seems to be promising in this respect. This relatively simple method
includes some effects of fluctuations and also works over the whole regime
of the parameter space. But the approximation involved in this scheme affects 
sometimes seriously the renormalization of the parameters (especially the 
off-diagonal ones like the single particle hopping amplitude). This, in turn, 
restricts the reliability of the numerical results produced by this method.
           
As far as the calculation of ground state energy is concerned,
the main difficulty of the method is the fact that by 
fixing a particular Boundary Conditions
(BC) on a block one may lose a number of states which contribute to the 
ground state of the whole lattice. 
This point is clearly
highlighted in 1D Tight Binding model in which the standard RG
fails for some type of BCs [8,9,10]. It must be noticed that
this difficulty is not removed by increasing the size of blocks.

In 1992 White [11] succeeded in greatly improving the
efficiency of the RG method by inventing  the Density Matrix Renormalization
Group (DMRG) scheme. In this approach, one embeds each
block into a larger(super) block and considers the block as a quantum
system in interaction with a reservoir (the rest of the superblock). 
The block can then be described by
a reduced density matrix, whose eigenkets with the highest eigenvalues
are used to construct the embedding operator ($T$) (see [12] and references therein),
which is the mathematical artifact which reduces the dimension of the Hilbert
space in each step of RG.
Although DMRG gives very accurate results compared with previous simple
RG schemes, its practical implementation is much more
difficult and time consuming.

Regarding the fermion systems an improved scheme to the standard RG 
was also proposed for free-fermion systems [13] which converges to 
the exact results of the ground state energy faster than the standard RG, 
but was not applied to interacting systems. 

In this paper we apply a modification of the standard RG scheme to a  
one dimensional interacting fermion system and obtain the ground state of the 
system with much higher accuracy compared with the standard RG [14]. 
For free-fermion model, our prescription yields much closer results to the 
exact values compared with those obtained from the standard RG and those obtained
in Ref.[13], we also obtain the density-density correlation function, 
which is in a rather good agreement with the exact one.

Our method, although does not yield as
much accuracy as in DMRG, is much simpler practically, so that one can easily
implement it on a personal computer. The required time is less than one second
for a Pentium machine. 
The results are compared  in table-1,  table-2 and Fig.3.
\section{ The Model }

We will consider the following 1D Hamiltonian, for spinless fermions 
\be
H=t\sum_{i}(\cd{i}\c{i+1}+\cd{i+1}\c{i})+G\sum_{i}(\n{i}\n{i+1})-\mu\sum_{i}\n{i}  \hspace{0.2cm},
\ee
where $\cd{i}$ and $\c{i}$ are the usual creation and annihilation fermion
operators on the $i$-th site and $\n{i}=\cd{i}\c{i}$ is the fermion number operator. 
When $G=0$ (free fermion case), $t$ represents the hopping 
parameter and $\mu$ is the site energy while
$G$ is a positive electron-electron repulsion between
electrons on neighbouring sites. We will consider 
the electron-hole symmetric case where there is an average 
of $\frac{1}{2}$ electron per sites (half-filled case). In this case the 
Fermi energy $\mu$ is conveniently taken equal to $G$ and, adding a constant,
the Hamiltonian is rewritten as follows 
\be 
H=t\sum_{i}(\cd{i}\c{i+1}+\cd{i+1}\c{i})+G\sum_{i}(\n{i}-\frac{1}{2})(\n{i+1}-\frac{1}{2})   \hspace{0.2cm}.
\ee

It is clear that the two terms in (2) play opposite roles. We can distinguish 
two different regimes :($i$) when $\frac{G}{2t}< 1$ the system is conducting
and $<\n{i}>=\frac{1}{2}$ on each site and $(ii)$ when $\frac{G}{2t}> 1$
the system organizes itself in two sub-lattices $A$ and $B$ such that $<\n{i}>\simeq0,
i \in A $ and $<\n{i}>\simeq1, i \in B $.

Our aim is to calculate the ground state energy of this fermionic system in the
conducting regime using a modified RG prescription. In section (3) we apply
the method to free-fermions ($G=0$) and in section (4) we will consider the
interacting case ($G\ne0$).

\section{ Modified RG for Free-Fermions }
In the free-fermion case the Hamiltonian will reduce to the following simple form 
\be
H=t\sum_{i}(\cd{i}\c{i+1}+\cd{i+1}\c{i}) \hspace{0.2cm},
\ee
where the hopping parameter $t$ is the only one that will be renormalized.

The decomposition of the lattice into isolated blocks and superblocks
with an odd number of sites per each block is shown in Fig.1.
In this case the size of blocks ($n_{B}$) and superblocks ($n_{SB}$) are taken 
to be 3 and 5 respectively for simplicity. Therefore the block Hamiltonian 
$h^{B}$ and  the superblock Hamiltonian $h^{SB}$ are 
\be
h^{B}=t(\cd{2}\c{3}+\cd{3}\c{2}+\cd{3}\c{4}+\cd{4}\c{3})  \hspace{0.2cm},
\ee
\be
h^{SB}=t(\cd{1}\c{2}+\cd{2}\c{1})+h^{B}+t(\cd{4}\c{5}+\cd{5}\c{4})  \hspace{0.2cm}.
\ee
We express our prescription in the following steps : 

{\bf step-1}) The superblock Hamiltonian ($h^{SB}$) is diagonalized exactly.
The doubly degenerate ground state is in the subspaces of particle number
$\nu=2$ and 3 ,
\be
h^{SB}\ket{\nu=2 (3)}_{SB}=\ep_{o}^{SB}(t)\ket{\nu=2 (3)}_{SB} \hspace{0.2cm},
\ee  
where $$\ep_{o}^{SB}(t)=(-2.73246)t \hspace{0.2cm}, $$ is the ground state
energy of the 5 sites superblock. Note that
the structure of the low lying levels of the superblock depends strongly on the
parity of $n_{SB}$. For $n_{SB}$ even, the ground state is a singlet corresponding
to the ground state of the subspace $\nu=\frac{n_{SB}}{2}$, while for $n_{SB}$
odd the ground state is a doublet corresponding to the electron-hole
degenerate ground state of the two subspaces $\nu=\frac{n_{SB}\pm 1}{2}$.
Here it is essential to restrict ourselves to the case $n_{SB}$ odd,
if we want, by keeping only two levels that the Hamiltonian conserves its
initial form.

{\bf step-2}) In this step in view of the half-filled property of the system
we project the degenerate ground state 
of the superblock ($\ket{\nu=2 (3)}_{SB}$) onto the $\nu=1$ and 2 subspaces
of the block Hilbert space respectively.{\it It is this step 
which effectively smoothes 
out the sharp effect of the boundary conditions
by immersing the block into a superblock, i.e.   
the projected state is not restricted by any
particular boundary conditions of an isolated block (open, periodic,...),
as compared with the standard RG, but takes into account to some extent 
the quantum fluctuation of the rest of the lattice.} 
Moreover by emphasizing the half-filled property, our RG prescription
leads to a form-invariant Hamiltonian. The resulting
normalized state of this projection is named $\ket{\nu=1 (2)}_{B}$. 

{\bf step-3)} The embedding operator is now constructed  
in the following form 

\be
T_{I}=(\ket{\nu=1}_{B}\bra{0})+(\ket{\nu=2}_{B}\bra{1}) \hspace{0.2cm},
\ee \\
where $\ket{0}$ and $\ket{1}$ are the renamed base kets of the 
effective Hamiltonian Hilbert space.

Having the form of the embedding operator, one can calculate the
projection of any operator onto the effective Hilbert space [12], 
(labelling the sites in the block 1, 2, 3 from left to right) we will have 
\bea
T_{I}^{\dagger} \c{jI}  T_{I}=\lambda \c{I}' \hspace{2cm} j=1,3  \hspace{0.2cm}, \nonumber \\
T_{I}^{\dagger} \cd{jI} T_{I}=\lambda \cd{I}' \hspace{2cm} j=1,3 \hspace{0.2cm},
\eea
where $\c{I}'$ and $\cd{I}'$ act as fermion operators in the new Hilbert space,
and $$\lambda^{2}=0.47999 \hspace{0.2cm}, $$ 
(a similar equation for $j=2$ can be obtained, but this is not required here).

Using Fig.2 and the labelling mentioned above for sites in each block, 
one writes the interaction between blocks in the following form 
\be
h_{I,J}^{BB}=t(\cd{3I}\c{1J}+\cd{1J}\c{3I}) \hspace{0.2cm}.
\ee
Then the effective Hamiltonian between the new sites is 
\be
h_{I,J}^{eff}=T_{J}^{\dag}T_{I}^{\dag}h_{I,J}^{BB}T_{I}T_{J} \hspace{0.2cm},
\ee    
where I and J are two neighbouring blocks, while $T_{I}$ and $T_{J}$ are
their corresponding embedding operators. This leads to
\be
h_{I,J}^{eff}=t'(\cd{I}'\c{J}'+\cd{J}'\c{I}')  \hspace{0.2cm},
\ee
where
\be   t'=\lambda^{2} t  \hspace{0.2cm},  \ee
and $c'_{I(J)}$ and $c'^{\dagger}_{I(J)}$ are fermion operators for the new sites.
This equation determines the renormalization of the coupling constant.

{\bf step-4)} The crudest estimate for the ground state energy is just the sum of 
ground state energies of all the blocks or superblocks. Thus,
\be
(\frac{E}{N})^{(0)}=\frac{1}{5}\ep_{o}^{SB}(t) \hspace{0.2cm},
\ee
or
\be
(\frac{E}{N})^{(0)}=\frac{1}{3}\ep_{o}^{B}(t)  \hspace{0.2cm},
\ee 
where $(\frac{E}{N})^{(0)}$ is the ground state energy per site of the whole system
in the zeroth order approximation. Eq.(14) in fact defines 
$\ep_{o}^{B}$ as 
\be 
\ep_{o}^{B}=\frac{3}{5}\ep_{o}^{SB} \hspace{0.2cm}.
\ee  

However this estimate clearly neglects the contribution to the energy from 
interaction between blocks. To take this missing contribution into account,
we add to $(\frac{E}{N})^{(0)}$ the ground state energy of the new lattice, and obtain 
\be
(\frac{E}{N})^{(1)}=\frac{1}{3}\ep_{o}^{B}(t)+\frac{1}{9}\ep_{o}^{B}(t') \hspace{0.2cm}.
\ee 
Iterating this procedure we finally obtain 
\be
\frac{E}{N}=\sum_{n=0}^{\infty}\frac{1}{3^{n+1}}\ep_{o}^{B}(t^{(n)}) \hspace{0.2cm},
\ee
or                                          
\be
\frac{E}{N}=\sum_{n=0}^{\infty}\frac{1}{3^{n+1}}(\frac{3}{5}\ep_{o}^{SB}(t^{(n)}) \hspace{0.2cm},
\ee 
where $t^{(0)}=t$ and $t^{(n)}$ is the renormalized coupling 
constant after $n$ steps of RG.
In fact, the above argument describes the physical interpretation of the formal 
limiting process defined in [12] in which the ground state energy is obtained as 
$$ E=lim _{n\rightarrow {\infty} } {T^{\dagger} }^n  H T^{n} \hspace{0.2cm}. $$
However in the standard RG the term $\ep_o^B (t) $ is the result of 
diagonalization of $ h^B $ , i.e., $T^{\dagger} h^B T= \ep_o^B $ , whereas in our
method $\ep_o^B $ is just another expression for $ \frac {n_{B}}{n_{SB}}\ep_o^{SB} $ where 
$\ep_{o}^{SB}$ is the eigenvalue of $ h ^{SB} $. 
The results for the ground state energy per site, obtained by this modified RG method
are collected in table-1. The results are compared both with the exact results 
and with those obtained from the standard RG [14]. It is  
seen that our modified RG scheme yields closer results to the exact one  
than the standard RG method.

We would like to comment on the non-variational nature of our
results. This is caused by our crude estimate for the ground state energy
of each block in terms of that of the superblock ($e_{o}^{B}=\frac{3}{5}e_{o}^{SB}$).
However one should note that this behaviour is not unusual in RG treatment
of lattice systems. For example in standard RG at second order perturbation
of free-fermion model [2] and isotropic Heisenberg model [16], 
one obtains non-variational results. We have also
obtained the ground state energy in this case
for other $\frac{n_{B}}{n_{SB}}$ ratios, namely for the ratios $\frac{5}{7}$,
$\frac{7}{9}$, $\frac{9}{11}$ and $\frac{11}{13}$,
the results are collected in table-1.
Note that our results in these cases
are above the exact values of ground state energy. This indicates that
the non-variational result in the $\frac{3}{5}$ case is due to the very small
block and superblock sizes. Therefore we expect that when the sizes 
of block and superblock are
not very small, our method must indeed reproduces the variational results.
Our data in table-1 show more accuracy compared with 
the standard RG results for any
values of the block size. These data also show that the result converges to the 
exact one by increasing the $\frac{n_{B}}{n_{SB}}$ ratios.
 
To have a better insight about the effective Hamiltonian obtained by this modified 
RG, one can look at the nature of the constructed wavefunction. In this 
respect we compute the density-density correlation function 
($\langle n_i n_{i+r} \rangle$) in terms of the distance $r$. In order to do so by
our RG prescription with a block size of $n_B = 3$, we consider 
a chain of length $3^N$ and let N go to infinity. In this calculation
the ground state of the whole system $\ket{0}$ is replaced by $T \ket{0'}$
where $\ket{0'}$ is the ground state of the effective Hilbert space. Then
we compute $g(r) =\langle n_i n_{i+r} \rangle$ for any value of $r$.
Our results are plotted in Fig.3 and compared with the exact results [17].
Our results show a very good qualitative agreement with the exact results while
the differences in the small $r$ is due the finite size effect of the 
exact results as mentioned in ref [17]. Although this good agreement 
is not specially due to our prescription and it can be obtained from 
other different schemes [2][3][4], this shows that the nature of the
constructed wavefunction does not change in our scheme and the obtained 
effective Hamiltonian is a good approximate one.

\section{Interacting Fermion system}
The system under consideration is an interacting fermion system 
with the Hamiltonian
defined in (2). Our RG prescription in this case is the same as in the 
free-fermion case. With $n_{B}=3$ and $n_{SB}=5$ the block and superblock 
Hamiltonians are (see Fig.1)
\be
h^{B}=t(\cd{2}\c{3}+\cd{3}\c{2}+\cd{3}\c{4}+\cd{4}\c{3}) 
+G(\n{2}-\frac{1}{2})(\n{3}-\frac{1}{2})+G(\n{3}-\frac{1}{2})(\n{4}-\frac{1}{2}) \hspace{0.2cm},
\ee
\be
h^{SB}=t(\cd{1}\c{2}+\cd{2}\c{1})+G(\n{1}-\frac{1}{2})(\n{2}-\frac{1}{2})+
h^{B}+t(\cd{4}\c{5}+\cd{5}\c{4})+G(\n{4}-\frac{1}{2})(\n{5}-\frac{1}{2}) \hspace{0.2cm},
\ee
Following the first three steps of our RG prescription we arrive at 
the effective operators 
\bea
T_{I}^{\dagger} \c{jI}  T_{I}=\eta \c{I}' \hspace{2cm} j=1,3  \hspace{0.2cm}, \nonumber \\
T_{I}^{\dagger} \cd{jI} T_{I}=\eta \cd{I}' \hspace{2cm} j=1,3 \hspace{0.2cm},
\eea
where $\eta$ is calculated to be 
\be
\eta=\frac{2\a\b}{2\a^2+\b^2}    \hspace{0.2cm}.
\ee
Here
\bea
\a=\ep^4+(1-g^2)\ep^2+g\ep-2  \hspace{0.2cm}, \nonumber  \\
\b=2(\ep-g)(2\ep^2+g\ep-2)    \hspace{0.2cm}, \nonumber  \\
\ep=\frac{1}{t}\ep_{o}^{SB}(t,G) \hspace{1cm}, 
\hspace{1cm} g=\frac{G}{2t} \hspace{0.2cm},
\eea
and as before $\ep_{o}^{SB}(t,G)$ is the ground state energy of $h^{SB}$ which
is some function of $t$ and $G$. Following Fig.2 the interaction between
blocks I and J is 
\be
h_{I,J}^{BB}=t(\cd{3I}\c{1J}+\cd{1J}\c{3I})+G(\n{3I}-\frac{1}{2})(\n{1J}-\frac{1}{2}) \hspace{0.2cm}.
\ee
Then the interaction between the new sites is given by
\be
h_{I,J}^{eff}=T_{J}^{\dag}T_{I}^{\dag}h_{I,J}^{BB}T_{I}T_{J} \hspace{0.2cm},
\ee    
where $T_{I}$ and $T_{J}$ are
the embedding operators of blocks I and J. In this way we arrive at 
\be
h_{I,J}^{eff}=t'(\cd{I}\c{J}+\cd{J}\c{I})+G'((\n{I}-\frac{1}{2})(\n{J}-\frac{1}{2})) \hspace{0.2cm},
\ee
where $t'$ and $G'$ are the renormalized coupling constants given by 
\bea
t'=\eta^{2} t \hspace{0.2cm},   \cr
G'=\gamma^{2} G  \hspace{0.2cm},
\eea
and 
\be
\gamma=\frac{\b^2}{2\a^2+\b^2} \hspace{0.2cm}.
\ee
After performing step-4 in the preceding section the ground state energy
of interacting fermion system will be obtained which is given in 
table-2. These data clearly show that the modified RG method yields 
much better results than the standard RG method.
The non-variational nature of these results is due to the same reason 
which was described in the free-fermion case. Moreover, these results will
be variational, if one takes larger $n_B$ and $n_{SB}$.

\section{Conclusions}

In  this letter we have introduced a version of RG scheme and applied
it to the free and interacting fermion systems. In this modified scheme,
we can take care of the effect of boundary conditions in a simple way. 
We have calculated the ground state energy of the free and interacting fermion 
model for different values of the block size. All of these results show more
accuracy than the standard RG and one can see the convergence of these results 
to the exact ones. In the free case we have also calculated 
the density-density correlation function whose good agreement with the exact results
verifies that the renormalized Hamiltonian is a 
good approximate Hamiltonian for the low energy spectrum of the original one.
Using the renormalization of any operator, one can also compute the 
correlation function for any value of $g(=\frac{G}{2t})$ in 
the interacting case.
This method can also be applied to spin systems [18].
One may hope that for a first 
estimate of many properties of these systems, our method or its improvements
(or adaptation for other lattice systems) would give quite good results. 
In those cases where application of the more sophisticated DMRG method 
seems to be hopeless, like the two dimensional Heisenberg Anti-ferromagnet,
we hope our method to give good approximate results. Work in 
this direction is in progress.

\section{Acknowledgement}
A. L. would like to thank M. R. Rahimi-Tabar for useful discussions, 
A. Shojaie  for his useful comments on
numerical computations, N. Heydari, A. Mostafazadeh for careful study of
the manuscript and Mrs. M. Shafaati for her help in editing the manuscript.

\newpage
{\Large \bf Figure Captions :}\\

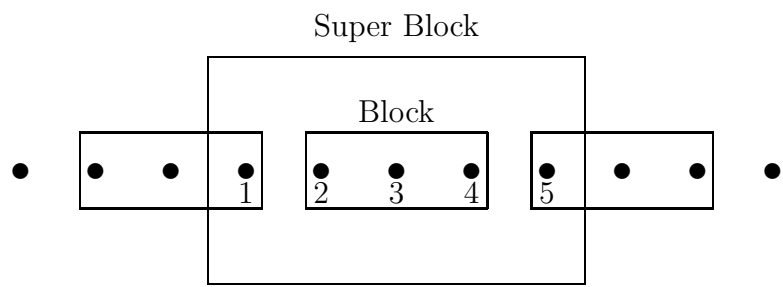
{\bf Fig.1)} Block and Super Block for the lattice chain.

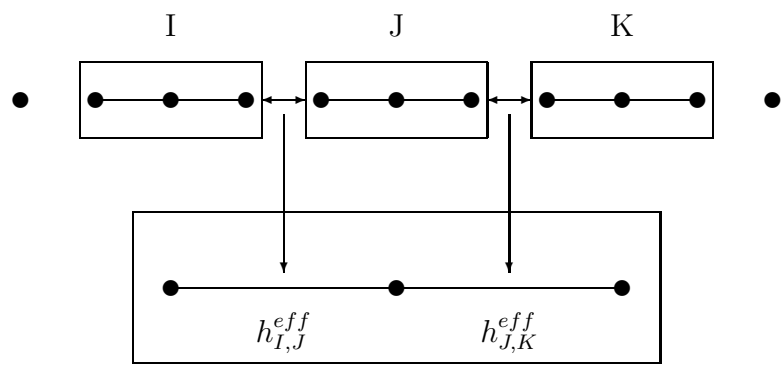
{\bf Fig.2)} The decomposition of the lattice into isolated blocks and the
consideration of 
neglected bonds as the effective interactions in the new Hilbert space.

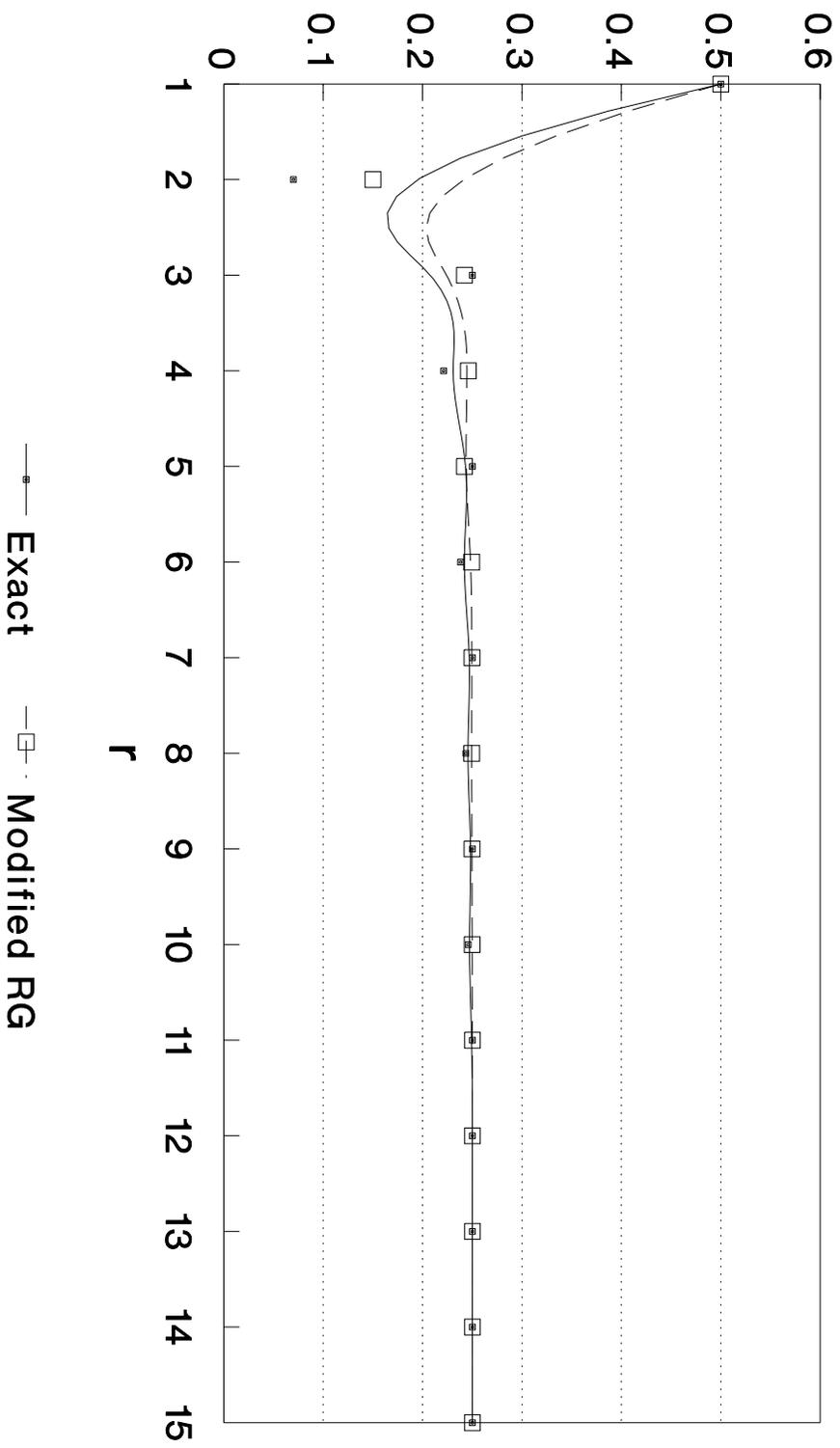
{\bf Fig.3)} Density-density correlation function $g(r)$ versus separation
$r = j-i$ between sites $i$ and $j$ for free-fermion chain.

\newpage
{\Large \bf Tables}\\
Table-1.Ground state energy per site of standard RG, exact results 
and modified RG for an infinite free-fermion chain for different values
of block size.
$$
\begin{array}{|c|c|c|c|}\hline
Exact \,\,result & Standard \,\,RG & Modified \,\,RG & Type \,\,of \,\,SB \,\,and \,\,B\\ \hline
-0.63662 (\frac{-2}{\pi}) & n_{B}=3, -0.565686 & -0.65058 & n_{SB}=5, n_{B}=3 \\ \hline
               & n_{B}=5, -0.5854 & -0.62256 & n_{SB}=7, n_{B}=5 \\ \hline
               & n_{B}=7, -0.5966 & -0.6193  & n_{SB}=9, n_{B}=7  \\ \hline
               & n_{B}=9, -0.6038 & -0.6196  & n_{SB}=11, n_{B}=9 \\ \hline
               & n_{B}=11, -0.6088 & -0.6206 & n_{SB}=13, n_{B}=11 \\ \hline       
\end{array}
$$
Table-2.Ground state energy per site of standard RG, exact results 
and modified RG for an infinite interacting fermion chain.
$$
\begin{array}{|c|c|c|c|c|c|c|}\hline
g(=\frac{G}{2t})&0&0.1&0.2&0.3&0.4&0.5 \\ \hline
(\frac{E}{tN})_{standard \,\,RG}&-0.565686&-0.584168&-0.603266&-0.623008&-0.643424&-0.664549  \\ \hline
(\frac{E}{tN})_{exact}      &-0.636620&-0.657384&-0.679129&-0.701826&-0.725454&-0.750000  \\ \hline
(\frac{E}{tN})_{modified \,\,RG}&-0.650583&-0.671674&-0.692069&-0.714005&-0.736275&-0.758952  \\ \hline
\end{array}
$$
$$
\begin{array}{|c|c|c|c|c|}\hline
0.6&0.7&0.8&0.9&1\\ \hline
-0.686424&-0.709098&-0.732630&-0.757097&-0.782609  \\ \hline  
-0.775452&-0.801804&-0.829058&-0.857217&-0.886294  \\ \hline  
-0.782419&-0.806508&-0.830934&-0.855876&-0.881525  \\ \hline   
\end{array}
$$

\newpage
\unitlength=1.00mm
\special{em:linewidth 0.4pt}
\linethickness{0.4pt}
\begin{center}  
\begin{picture}(151.00,109.00)
\put(30.00,80.00){\circle*{2.00}}
\put(40.00,80.00){\circle*{2.00}}
\put(50.00,80.00){\circle*{2.00}}
\put(60.00,80.00){\circle*{2.00}}
\put(70.00,80.00){\circle*{2.00}}
\put(80.00,80.00){\circle*{2.00}}
\put(90.00,80.00){\circle*{2.00}}
\put(100.00,80.00){\circle*{2.00}}
\put(110.00,80.00){\circle*{2.00}}
\put(120.00,80.00){\circle*{2.00}}
\put(130.00,80.00){\circle*{2.00}}
\put(38.00,75.00){\framebox(24.00,10.00)[cc]{}}
\put(68.00,75.00){\framebox(24.00,10.00)[cc]{}}
\put(98.00,75.00){\framebox(24.00,10.00)[cc]{}}
\put(60.00,77.00){\makebox(0,0)[cc]{1}}
\put(70.00,77.00){\makebox(0,0)[cc]{2}}
\put(80.00,77.00){\makebox(0,0)[cc]{3}}
\put(90.00,77.00){\makebox(0,0)[cc]{4}}
\put(100.00,77.00){\makebox(0,0)[cc]{5}}
\put(80.00,99.00){\makebox(0,0)[cc]{Super Block}}
\put(80.00,88.00){\makebox(0,0)[cc]{Block}}
\put(80.00,45.00){\makebox(0,0)[cc]{Figure 1}}
\put(55.00,65.00){\framebox(50.00,30.00)[cc]{}}
\end{picture}

\begin{picture}(151.00,100.00)
\put(30.00,80.00){\circle*{2.00}}
\put(40.00,80.00){\circle*{2.00}}
\put(50.00,80.00){\circle*{2.00}}
\put(60.00,80.00){\circle*{2.00}}
\put(70.00,80.00){\circle*{2.00}}
\put(80.00,80.00){\circle*{2.00}}
\put(90.00,80.00){\circle*{2.00}}
\put(100.00,80.00){\circle*{2.00}}
\put(110.00,80.00){\circle*{2.00}}
\put(120.00,80.00){\circle*{2.00}}
\put(130.00,80.00){\circle*{2.00}}
\put(38.00,75.00){\framebox(24.00,10.00)[cc]{}}
\put(68.00,75.00){\framebox(24.00,10.00)[cc]{}}
\put(98.00,75.00){\framebox(24.00,10.00)[cc]{}}
\put(62.00,80.00){\vector(1,0){6.00}}
\put(92.00,80.00){\vector(1,0){6.00}}
\put(97.00,80.00){\vector(-1,0){5.00}}
\put(67.00,80.00){\vector(-1,0){5.00}}
\put(40.00,80.00){\line(1,0){20.00}}
\put(70.00,80.00){\line(1,0){20.00}}
\put(100.00,80.00){\line(1,0){20.00}}
\put(50.00,55.00){\circle*{2.00}}
\put(80.00,55.00){\circle*{2.00}}
\put(110.00,55.00){\circle*{2.00}}
\put(45.00,45.00){\framebox(70.00,20.00)[cc]{}}
\put(50.00,55.00){\line(1,0){60.00}}
\put(65.00,78.00){\vector(0,-1){21.00}}
\put(95.00,78.00){\vector(0,-1){21.00}}
\put(65.00,49.00){\makebox(0,0)[cc]{$h_{I,J}^{eff}$}}
\put(95.00,49.00){\makebox(0,0)[cc]{$h_{J,K}^{eff}$}}
\put(50.00,90.00){\makebox(0,0)[cc]{I}}
\put(80.00,90.00){\makebox(0,0)[cc]{J}}
\put(110.00,90.00){\makebox(0,0)[cc]{K}}
\put(80.00,25.00){\makebox(0,0)[cc]{Figure 2}}
\end{picture}
\end{center}


\newpage 
\begin{thebibliography}{99} 
\bibitem{1} K. G. Wilson, Rev. Mod. Phys. {\bf 47} (1975) 773.
\bibitem{2} P. Pfeuty, R. Jullien and K. L. Penson in : Real-Space Renormalization
, eds.T. W. Burkhardt and J. M. J. van Leeuwen (Springer, Berlin, 1982) ch.5.
\bibitem{3} J. E. Hirsch, Phys. Rev. {\bf B 22} (1980) 5259.
\bibitem{4} C. Dasgupta and P. Pfeuty, J. Phys. {\bf C 14} (1981) 717.
\bibitem{5} B. Bhattacharyya and S. Sil, Phys. Lett. {\bf A 180} (1993) 299;
J. Phys. Cond. Matt. {\bf 7} (1995) 6663; B. Bhattacharyya and G. K. Roy,
J. Phys. Cond. Matt. {\bf 7} (1995) 5537.
\bibitem{6} J. Perez-Conde and P. Pfeuty, Phys. Rev. {\bf B 47} (1993) 856.
\bibitem{7} J. Yi, L. Zhang and G. S. Canright, Phys. Rev. {\bf B 49} (1994) 15920.
\bibitem{8} S. R. White and R. M. Noak, Phys. Rev. Lett. {\bf 68} (1992) 3487.
\bibitem{9} M. A. Martin-Delgado and G. Sierra, Phys. Lett. {\bf B364} (1995) 41.
\bibitem{10} M. A. Martin-Delgado, J. Rodriguez-Laguna and G. Sierra, Nucl. Phys.
{\bf B473} (1996) 685.
\bibitem{11} S. R. White, Phys. Rev. Lett. {\bf 69} (1992) 2863; S. R. White,
Phys. Rev. {\bf B48} (1993) 10345.
\bibitem{12} M. A. Martin-Delgado and G. Sierra, Int. J. Mod. Phys. {\bf A11} (1996) 3145,
J. Gonzalez, M. A. Martin-Delgado, G. Sierra, A. H. Vozmediano in :
Quantum Electron Liquids and High-$T_{c}$ Superconductivity, Lecture Notes
in Physics, Monographs. Vol 38 (Springer, Berlin, 1995) ch.11.
\bibitem{13} B. Bhattacharyya and S. Sil, Phys. Lett. {\bf A 210} (1996) 129.
\bibitem{14} G. Spronken, R. Jullien and M. Avignon, Phys. Rev. {\bf B 24} (1981) 5356.
\bibitem{15} H. Bethe, Z. Physik {\bf 71} (1931) 205; R.Orbach, phys. Rev.
{\bf 112} (1958) 309.
\bibitem{16} M. A. Martin-Delgado, Proceedings of the El Escorial Summer School
1996 on Strongly Correlated Magnetic and Superconducting Systems; cond-mat/9610196.
\bibitem{17} U. Busch and K. A. Penson, Phys. Rev. {\bf B36} (1987) 9271.
\bibitem{18} V. Karimipour and A. Langari,"A Modified Quantum Renormalization 
Group for the {\bf xxz} Spin Chain", IPM-96-167; cond-mat/9611057.

\end{thebibliography}
\end{document}